# Multidimensional thermally-induced transformation of nest-structured complex Au-Fe nanoalloys towards equilibrium

Jacob Johny[a,‡], Oleg Prymak[b,‡], Marius Kamp[c], Florent Calvo[d], Se-Ho Kim[e], Anna Tymoczko[a], Ayman El-Zoka[e], Christoph Rehbock[a], Ulrich Schürmann[c], Baptiste Gault[e,f], Lorenz Kienle[c] and Stephan Barcikowski[a,*]

[a] Technical Chemistry I and Center for Nanointegration Duisburg-Essen (CENIDE), University Duisburg-Essen, Universitätstr. 7, 45141 Essen, Germany

[b] Inorganic Chemistry and Center for Nanointegration Duisburg-Essen (CENIDE), University of Duisburg-Essen, Universitätsstr. 7, 45141 Essen, Germany

[c] Institute for Materials Science, Synthesis and Real Structure, Kiel University, Kaiserstr. 2, 24143 Kiel, Germany

[d] University Grenoble Alpes and CNRS, LIPHY, F-38000 Grenoble, France

[e] Max-Planck-Institut für Eisenforschung GmbH, Max-Planck-Str. 1, 40237 Düsseldorf, Germany

[f] Department of Materials, Royal School of Mine, Imperial College London, Exhibition Road, London SW7 2AZ, United Kingdom

[‡] These authors contributed equally

# ABSTRACT

Bimetallic nanoparticles are often superior candidates for a wide range of technological and biomedical applications, thanks to their enhanced catalytic, optical, and magnetic properties, which are often better than their monometallic counterparts. Most of their properties strongly depend on their chemical composition, crystallographic structure, and phase distribution. However, little is known of how their crystal structure, on the nanoscale, transforms over time at elevated temperatures, even though this knowledge is highly relevant in case nanoparticles are used in, e.g., high-temperature catalysis. Au-Fe is a promising bimetallic system where the low-cost and magnetic Fe is combined with catalytically active and plasmonic Au. Here, we report on the *in situ* temporal evolution of the crystalline ordering in Au-Fe nanoparticles, obtained from a modern laser ablation in liquids synthesis. Our in-depth analysis, complemented by dedicated atomistic simulations, includes a detailed structural characterization by X-ray diffraction and transmission electron microscopy as well as atom probe tomography to reveal elemental distributions down to a single atom resolution. We show that the Au-Fe nanoparticles initially exhibit highly complex internal nested nanostructures with a wide range of compositions, phase distributions, and size-depended microstrains. The elevated temperature induces a diffusion-controlled recrystallization and phase merging, resulting in the formation of a single face-centered-cubic ultrastructure in contact with a body-centered cubic phase, which demonstrates the metastability of these structures. Uncovering these unique nanostructures with nested features could be highly attractive from a fundamental viewpoint as they could give further insights into the nanoparticle formation mechanism under non-equilibrium conditions. Furthermore, the *in situ* evaluation of crystal structure changes upon heating is potentially relevant for high-temperature process utilization of bimetallic nanoparticles, e.g., during catalysis.

# INTRODUCTION

Heterogeneous catalysis is one of the fields where bimetallic nanoparticles (NPs) have attracted much attention due to their enhanced physicochemical properties compared to the monometallic counterparts. [1-3] Many industrially relevant catalytic operations, including CO oxidation, ethylene hydrogenation, and partial oxidation, however, take place at temperatures above 300 °C. [4,5] Such thermal treatments will lead to enhanced atomic mobility in and between the metallic constituents by overcoming the diffusion activation energy barrier, [6] thereby influencing the resulting properties. This makes it of great interest to investigate the *in situ* temporal evolution of the structure, composition, and phase inhomogeneities of the bimetallic NPs at catalytically relevant temperature regimes to further fine-tune their performance. [7-10] Au and Fe-based nanostructures have already shown their potential as catalysts in many reactions, including oxygen evolution and $CO_2$ reduction. [11,12] They also possess additional advantages of reducing the abundance of high-cost noble metals together with chemical inertness, biocompatibility, and visible range plasmonic properties of Au and high saturation magnetization of Fe. [13-16] Besides catalysis, they have been found to be promising candidates for other applications as well, including multimodal magnetic resonance imaging (MRI), data storage, and biomedicine owing to the synergistic effects resulting from the combination of both elements. [12,17-20] The Au-Fe system has been extensively investigated also due to the equilibrium immiscible behavior and different crystal structures (face-centered cubic, fcc, for Au, and body-centered cubic, bcc, for Fe) of the constituting elements. [21-23] Even though Au-Fe alloy NPs have been obtained through various synthetic approaches to date, [24-27] the high sensitivity of Fe towards oxidation is often a major challenge at the nanoscale. However, this issue can be overcome by protecting the magnetic Fe core inside an inert thin Au shell while simultaneously preserving the properties of both constituents. [28] Moreover, among different bimetallic nanostructures, core-shell (CS) configurations are a relatively new class where different functions are integrated into one in a controlled fashion along with optimized morphologies and compositions. [29] Many different strategies have thus been reported for producing CS Fe-Au NPs with simultaneous evaluation of composition and size-dependent properties or structures. [30-32] While Rabkin *et al.* demonstrated the synthesis of CS Fe-Au NPs by solid-state dewetting of thin Fe/Au bilayer exploiting the segregation of equilibrium Au on Fe particle surfaces and interfaces, [33] superparamagnetic Fe-Au CS NPs were synthesized by a laser-assisted chemical synthesis where Fe NPs prepared by wet chemistry were irradiated by a laser in the presence of Au powder in the liquid phase. [34]

Out of different synthesis methods for producing Au-Fe bimetallic NPs, [35,36] pulsed laser ablation in liquid (PLAL) is a reliable and environment-friendly approach. [37] The ability to produce ligand-free particles for direct use in biomedical applications, as well as the capability to generate non-equilibrium complex structures, make PLAL a highly interesting method for the generation of Au-Fe bimetallic NPs. [37,38] Laser ablation of bulk Au-Fe targets in liquids was shown to generate either solid solution (SS) or CS particles. [15,30,39,40] For instance, $Au_{89}Fe_{11}$ SS particles were synthesized by Amendola *et al.* by PLAL, where both magnetic and plasmonic properties co-existed. [15] The synthesis of Fe-Au CS NPs by PLAL was first reported by Wagener *et al.* by ablating an Au-Fe target in acetone or methyl methacrylate. [40] The strong dependence of either the CS or SS formation on the NP diameter and structure and composition of the Au-Fe alloy target was later shown by Tymoczko *et al.* for laser-generated Au-Fe NPs. [30,39] The bimetallic particles, in that case, presented substituted fcc or bcc phases where the bcc phase was dominant for higher molar Fe contents. This also underlines the significance of selecting the laser-generated alloy NPs for various post studies due to their highly complex internal structures. Since the physicochemical properties of Au-Fe NPs potentially depend on the inhomogeneous phase distribution, [20,41,42] a detailed analysis of such irregularities at the near-atomic scale and their temporal evolution is highly relevant for fundamental studies of nanoparticle formation far from room temperature thermodynamic equilibrium. Furthermore, this kind of metastable structural complexity may also play a significant role in their high-temperature catalytic performance. [9,10,20,43,44]

Based on these challenging aspects of bimetallic nanosystems from both fundamental and application points of views, the main focus of the present study is to (1) identify the minuscule inhomogeneities inside the NPs using scanning transmission electron microscopy (STEM) and atom probe tomography (APT) and (2) follow their structural evolution *in situ* by X-ray diffraction (XRD) and STEM at catalytically-relevant temperature regimes. Here, laser-generated Au-Fe bimetallic NPs present multiple fcc ($Au_{80}Fe_{20}$, $Au_{50}Fe_{50}$) and fcc/bcc ($Au_{20}Fe_{80}$) phases differing in their local composition, lattice strain, and crystallite size. Application of sufficient temperature above ~500 °C, where the activation energy barrier starts to decrease, initiates a transformation of the metastable Au-Fe NPs towards thermodynamic equilibrium by facilitating the mobility of the metal atoms (diffusion). This is also supported by independent atomistic simulations of the annealing of Au-Fe assemblies. Monitoring structural evolution, recrystallization, and relaxation enables the detection of the temperature regions, where such inhomogeneous formations (nests) exist in Au-Fe NPs before the thermally induced transformations set in.

**RESULTS AND DISCUSSION**

Non-equilibrium Au-Fe NPs of varying compositions ($Au_{80}Fe_{20}$, $Au_{50}Fe_{50}$, and $Au_{20}Fe_{80}$) were obtained through a liquid-based laser ablation method. The detailed crystal structure analysis was performed on *in situ* diffractograms collected from the NPs (Fig. S1 in the Supporting Information, SI) employing Rietveld refinement (representatively shown in Fig. S2 in SI) to determine phase fractions and atomic substitutions. Special attention was paid to individual shoulder reflections that emerged due to multiple fcc phases, which originated from different local compositions and crystallite sizes inside Au-Fe NPs (see Table S1 in SI). Monometallic Au and Fe NPs were also produced and investigated in the same manner for comparison (Fig. S3 and Table S2 in SI). In the case of $Au_{80}Fe_{20}$, two fcc phases were detected exhibiting larger (~80 nm) and smaller (~7 nm) crystallites labeled as fcc1 and fcc2, respectively (Fig. 1A). The existence of multiple fcc phases might indicate the presence of complex structures inside Au-Fe NPs produced by PLAL.

 Crystallite sizes of both fcc phases increase exponentially with increasing temperature until above 600 °C, and then both fcc phases merge, forming a single FCC phase at 700 °C with a crystallite size of ~107 nm as shown in the lattice parameter plots of both fcc phases vs. temperature (Fig. 1A, B).

It should be noted here that the small (fcc1, fcc2, etc.) and capital (FCC and BCC) letter style adopted throughout the text is to distinguish between differently substituted multiple and single phases, respectively. Typically, the Scherrer equation [45] is employed for crystallite size calculations down to 100 nm. [46] However, depending on the accuracy of the instrumental characterization and resolution of the diffractometer used for the Rietveld refinement (here $FWHM_{(inst)}=0.05°$), the crystallite size from the Scherrer equation can be calculated up to 200 nm as it was shown in the theoretical work of Miranda *et al.* [47] and experimentally applied in the work of Helmlinger *et al.* [48] Thus, the crystallite sizes below 200 nm estimated here are reliable. The increase in the fcc1 and fcc2 lattice parameters with temperature is attributed mostly to (I) the thermal expansion of the lattice, but also to (II) changes in the Au-Fe substitution ratio caused by the enhanced diffusion, which is also accompanied by (III) a decrease in the lattice strain. The initial fcc1 and fcc2 lattice parameters (4.029 and 4.05 Å) correspond to average atomic compositions of $Au_{88}Fe_{12}$ and $Au_{92}Fe_{08}$, which are close to the nominal composition $Au_{80}Fe_{20}$. For both fcc phases up to 700 °C, the corresponding lattice parameters are smaller in comparison with the lattice parameter of pure fcc Au NPs at the same temperature 700 °C (4.123 Å, Fig. S3 and Table S2 in SI, also in a good agreement with Ref. [49]), confirming partial substitution of larger Au (atomic radius 0.144 nm) [50] by smaller Fe (atomic radius 0.126 nm) [51] atoms in the fcc lattice. As shown in Fig. 1B, above 700 °C, only one FCC phase is detectable (4.078 Å), pointing to the stabilized atomic exchange between fcc1 and fcc2 caused by the facilitated diffusion. It should be noted that thermal expansion of the lattice increases the interatomic distances in favor of diffusion, providing needed room for the movement of the atoms. [52] This thermally activated process is also in agreement with the self-diffusion coefficients for pure Au and Fe [53, 54], which exponentially start to increase at ~600 °C (Fig. S4a in SI). During this interaction, Au atoms would migrate from fcc2 ($Au_{92}Fe_{08}$) to fcc1 ($Au_{88}Fe_{12}$) with a simultaneous motion of Fe atoms from fcc1 to fcc2, thereby balancing the ratios in both and sacrificing the phase amount of fcc2 in favor of fcc1 to form a single fcc (Fig. S3 and Table S3 in SI). A further thermal expansion of fcc1 together with an almost unchanged lattice parameter of fcc2 before the formation of a single FCC underlines this concomitant atom movements within the fcc lattices (Fig. 1B).

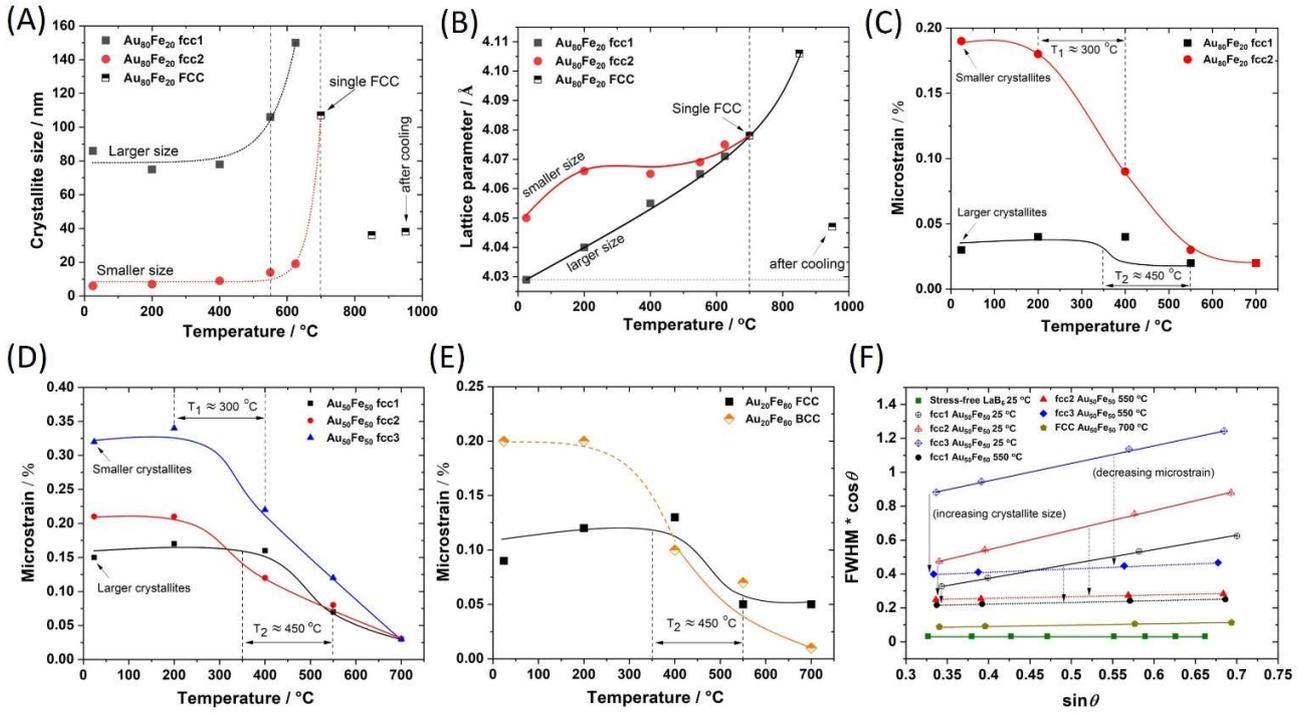

**Figure 1.** XRD derived (A) crystallite sizes, (B) lattice parameters and (C) microstrain of fcc1 and fcc2 phases for $Au_{80}Fe_{20}$ NPs (top) at varying temperatures, incl. after cooling down the NPs from 850 °C to room temperature. The microstrain (D) of fcc1, fcc2 and fcc3 for $Au_{50}Fe_{50}$ NPs and (E) of fcc and bcc for $Au_{20}Fe_{80}$ NPs (bottom). Note: FCC or BCC underlines a single phase. $T_1/T_2$ represents the approximate onset temperature of recrystallization in the corresponding phases pointing on a temperature-induced structure relaxation in all phases. (F) Representative Williamson-Hall (W-H) plots of different fccs for $Au_{50}Fe_{50}$ NPs showing decreasing lattice strain (line slopes) and increasing crystallite size (line intercepts) with increasing temperature.

The XRD patterns were also recorded at room temperature (RT) after cooling down the particles from 850 °C to confirm the merging trend of both fcc phases. After cooling down, the determined lattice parameter is 4.047 Å (Fig. 1B), a value which is located between the initial lattice parameters of fcc1 (4.029 Å) and fcc2 (4.050 Å), indicating the formation of one FCC structure with a more homogeneous atomic distribution within the fcc lattice. A decrease in the crystallite size of the formed single FCC lattice at 850 °C to approximately 36 nm, which was confirmed again for the cooled down sample (38 nm), can be attributed to the onset of evaporation of Au-Fe NPs (Fig. S1a in SI). [55, 56]

It is known that the broadening of XRD reflections is related to two physical effects, the change in the crystallite size or in the lattice strain. [57, 58] Therefore, to differentiate between them, microstrain was calculated for different phases as a function of temperature (Fig. 1 C-E) from the corresponding Williamson-Hall (W-H) plots (Fig. 1F). As can be seen in Fig. 1C, the fcc2 phase with smaller crystallites presents a significantly larger microstrain (0.19 %) in the lattice at RT compared to the fcc1 phase with larger crystallites (0.03 %) in the $Au_{80}Fe_{20}$ NPs. With increasing temperature, the microstrain decreased in both fcc phases, pointing towards started relaxation, but following different trends, e.g., earlier for smaller fcc2 ($T_1 \approx 300$ °C) and later for larger fcc1 crystallites ($T_2 \approx 450$ °C), attaining a relaxed state above 550 °C for both. This process might be accompanied by decreased atomic defects at higher temperatures quantified as microstrain. Such a temperature-induced relaxation was also observed by Prymak et al. [49] for small pure Ag and Au and alloyed Ag-Au NPs (with diameters of 35, 8, and 7 nm, respectively), where the process started even earlier at 200 °C for smaller Au and Ag-Au NPs and at 250 °C for larger Ag NPs.

As the diffusion-based interaction between different fcc phases in the $Au_{80}Fe_{20}$ NPs was noted by XRD, similar crystallographic studies were conducted on $Au_{50}Fe_{50}$ NPs complemented by *in situ* STEM analysis (Fig. 2A-E). The XRD data on $Au_{50}Fe_{50}$ NPs also verified the structural complexity of the NPs associated with different Au-Fe substitutions by detecting three fcc phases (consecutively labeled fcc1, fcc2, and fcc3) having distinct lattice parameters (Fig. 2D). The fcc1, fcc2, and fcc3 phases with average atomic compositions $Au_{74}Fe_{26}$, $Au_{82}Fe_{18}$, and $Au_{96}Fe_{04}$ exhibit crystallite sizes 23, 15, and 8 nm (Fig. 2E), respectively. Here it is important to note that according to the Au-Fe phase diagram [59], the $Au_{50}Fe_{50}$ NPs should also contain a bcc Fe phase, which in our case, is first detected at 700 °C for ~7 nm crystallites, increasing their size to 13 nm at 850 °C (see Fig. S1b and Table S1 in SI). The overestimated Au-rich compositions in the multiple fcc phases compared to the nominal $Au_{50}Fe_{50}$ suggests the existence of missing Fe in the bcc lattice, whose detection is challenging due to tiny crystallites. The XRD measurements for the $Au_{50}Fe_{50}$ NPs confirm that fcc1/fcc2/fcc3 phases with different Au-Fe substitution ratios transformed into a single FCC lattice at 700 °C (Fig. 2D), reaching a more homogeneous atomic distribution. At lower temperatures, the lattice parameters of all fcc phases increase due to thermal expansion of the unit cells as shown for the $Au_{80}Fe_{20}$ NPs as well but then follow an unusual decreasing tendency above 550 °C through facilitated diffusion. Normally, such atomic movements would follow a direction from higher to lower concertation of atoms, e.g., Fe from fcc1 ($Au_{74}Fe_{26}$) to fcc3 ($Au_{96}Fe_{04}$) and Au in the opposite direction, thereby increasing the lattice parameter of fcc1 as observed for the $Au_{80}Fe_{20}$ NPs. Since this is not in our observations, we hypothesize that the fcc lattices in the $Au_{50}Fe_{50}$ NPs also incorporated Fe from the Fe-bcc nests to decrease the lattice parameters in all fccs. After the formation of a single FCC and detection of BCC at 700 °C with crystallite sizes of about 97 nm and ~7nm, respectively (Fig. 2E), both phases underwent the normal thermal expansion upon further heating (Fig. 2D and Table S1 in SI). The untypical behavior of the fcc lattice parameters for $Au_{50}Fe_{50}$ NPs (Fig. 2D) is well reflected from the shifted reflection maxima in the diffractograms (see Fig. S1 in SI), which resemble the shape of a banana (called here as "banana effect"). After cool down, the FCC lattice parameter decreased to a=3.974 Å, a value smaller than the lattice parameter of FCC at 850 °C (a=3.999 Å) but still larger compared to that of fcc1 at RT (a=3.962 Å). The corresponding lattice parameter of the Fe-bcc phase decreased to a≈2.85 Å in comparison with the value at 850 °C (2.872 Å) following the same unit cell contraction phenomena. They simultaneously indicate a previously occurring diffusion between different phases (fcc-fcc and fcc-bcc). With the determined atomic compositions of FCC ($Au_{75}Fe_{25}$) and BCC (~$Fe_{100}$) and the FCC/BCC mass ratio (71/29, see Table S3 in SI) after cooling, a good agreement to the nominal composition is met (atomic Au/Fe=50/50 corresponds to weighted Au/Fe=78/22).

In view of the uncovered structural complexity in $Au_{80}Fe_{20}$ as well as $Au_{50}Fe_{50}$ compositions, the $Au_{50}Fe_{50}$ NPs were additionally subjected to *in situ* STEM at RT (Fig. 2A, incl. EDX line scan Fig. 2B), 550 and 700 °C (Fig. 2C) to visualize the complex internal structures with different Au-Fe substitutions inside the NPs and their thermally-induced transformation. Local elemental segregations (called as nests, Fig. 2A) inside the same NPs are clearly visible due to their distinguishable Z-contrasts and indicate that the NPs are constituted by highly inhomogeneous structures, as also previously reported for Au-Fe NPs by PLAL [20]. To resolve such nanometer-sized nests in the Au-Fe NPs by EDX, cross-sectional sample preparation of single NP by focus ion beam (FIB) technique was used (Fig. 2B). [60] The results agree with our expectations of temperature-induced recrystallization through enhanced elemental mixing as already assumed for $Au_{80}Fe_{20}$ NPs. At 700 °C, the $Au_{50}Fe_{50}$ NPs show an enhanced homogenous mixing over the entire individual NPs in comparison to that at RT (Fig. 2C), hence pointing to the unified single FCC phase detected in XRD. Comparable observations of the elemental mixing at elevated temperatures were also shown for Ag-Au alloyed NPs by STEM and *in situ* XRD. [49]

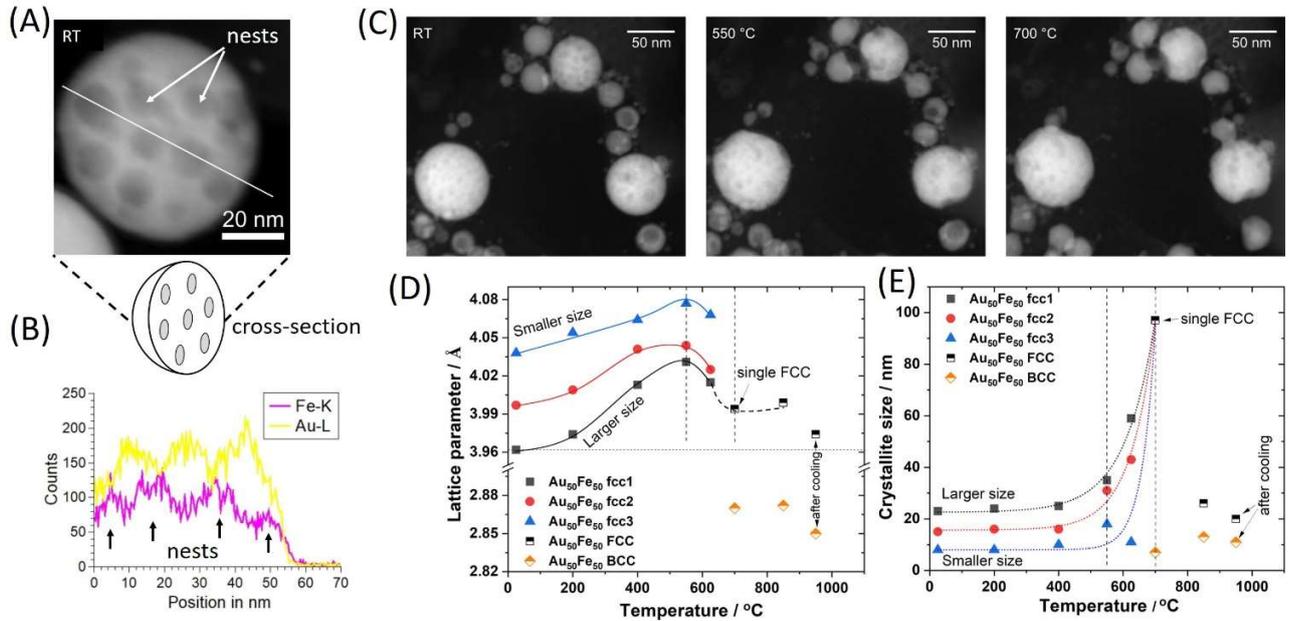

**Figure 2.** Representative STEM Z-contrast images of (A) a cross-sectional single $Au_{50}Fe_{50}$ NP at room temperature (RT) showing different contrast regions (darker Fe-rich and lighter Au-rich) as confirmed by the corresponding (B) EDX linescan and (C) *in situ* STEM images of multiple $Au_{50}Fe_{50}$ NPs at RT, 550 °C and 700 °C. Various Z-contrast formations inside NP caused by different Au-Fe substitutions are called as nests and marked by the arrows in (A) and (B). XRD derived (D) lattice parameters and (E) crystallite sizes of fcc1, fcc2, fcc3 and bcc phases for $Au_{50}Fe_{50}$ NPs. Note: FCC or BCC underlines a single phase.

The atomic exchange between different phases, as noted by *in situ* XRD and STEM, can be further facilitated by decreasing atomic defects, including grain boundaries. In the $Au_{50}Fe_{50}$ NPs, fcc1, fcc2, and fcc3 exhibit different size-depended microstrains (0.15, 0.21, and 0.32 %, respectively) at RT (Fig. 1D) as shown by distinguishable slopes in the W-H plots (Fig. 1F) following the same trend observed for $Au_{80}Fe_{20}$ NPs (higher microstrain for smaller crystallites). The detected lattice strains can be attributed again to an inhomogeneous distribution of Au and Fe atoms inside each fcc phase and limited mobility of the crystal defects especially accumulated on the grain boundaries inside the NPs. Swift cooling of the ablated species in PLAL [38, 61] could account for the formation of such complex internal structures and strongly influences the structure into which the atoms crystallize. [62] When the $Au_{50}Fe_{50}$ NPs are heated to 550 °C, the strained fcc phases transform into a more relaxed structure characterized by smaller slopes in the W-H plots in comparison with those at RT (Fig. 1F). Upon further heating to 700 °C, they display further structure relaxation, which is comparable to the stress-free $LaB_6$ reference. Interestingly, the fcc1 (larger crystallites) structure relaxation starts at a higher temperature (~450 °C) in comparison with the fcc2 and fcc3 phases (~300 °C), as shown in Fig. 1D. The found relaxation regions for fcc phases in the $Au_{50}Fe_{50}$ NPs agree with the trend observed for $Au_{80}Fe_{20}$ NPs and their dependence on the crystallite size. Moreover, our findings match the TEM observation of temperature-induced rearrangements within a

Fe−Au NP (~5 nm) [63] and diffusivity of Fe into Au-fcc and Au into Fe-bcc (see S4b in SI) [64, 65] with an onset temperature at ~400 °C. It should be noted that both $T_1$ and $T_2$ regions in the Au-Fe NPs are below the typical value for relaxation of the corresponding bulk materials (~0.6 melting point, [66, 67]), evidencing a size-dependent effect. Similar behavior was also observed in alloyed Ag-Au bulk at 720 °C in contrast to 200 °C for the 8 nm-sized Ag-Au NPs at the same composition. [49, 68] Along with the decrease in the microstrain, the Y-intercepts of the W-H plots corresponding to all fcc phases in the $Au_{50}Fe_{50}$ NPs also decrease (Fig. 1D) indicating an increase in the crystallite size with temperature. Thus, temperature-induced structure relaxation and recrystallization are verified in the $Au_{50}Fe_{50}$ NPs like in the $Au_{80}Fe_{20}$. Heating the NPs facilitates the internal movement of the Au and Fe atoms and leads to their enhanced homogeneous distributions (see also Fig. 2C).

To further address the question of the nest-constructed Au-Fe NPs at the near-atomic scale, our study was extended to Fe-rich composition analyzed by atom probe tomography (APT) in addition to the *in situ* XRD and STEM (Fig. 3A-G). In comparison to the Au-rich and intermediate Au-Fe, $Au_{20}Fe_{80}$ NPs contain single FCC and BCC phases, with the RT atomic compositions $Au_{75}Fe_{25}$ and $Au_{04}Fe_{96}$ having a mass ratio 54/46 (Table S3 in SI), respectively, as determined by XRD. Considering the Au/Fe substitution and phase amount, this together would correspond to an overall atomic ratio Au/Fe=25/75, which is close to the nominal $Au_{20}Fe_{80}$ composition. The temperature-induced thermal expansion of the FCC phase changes the behavior after 400 °C (Fig. 3A) starting to decrease (previously mentioned "banana" effect, see Fig. S1 in SI), so that diffusion plays a dominant role above this temperature. Concomitantly, the lattice parameter of BCC increases above 400 °C, following an opposite trend to the FCC phase. This underlines a co-dependent diffusion between FCC and BCC, more pronounced in the $Au_{20}Fe_{80}$ compared to the $Au_{50}Fe_{50}$ NPs, which exhibit bcc and dominant multiple fcc phases (Fig. 2D). The determined BCC lattice parameter in $Au_{20}Fe_{80}$ after cooling (2.872 Å) is smaller compared to the previous value at RT (2.888 Å) and closer to the parameter of monometallic Fe NPs (2. 856 Å, see Table S2 in SI). This further indicates that a few of the Au atoms incorporated in the BCC lattice have migrated into the FCC lattice, simultaneously contracting the BCC and expanding the FCC (from 3.975 Å to 3.983 Å) unit cells. Once heated to 700 °C, the BCC Fe-rich crystals of $Au_{20}Fe_{80}$ NPs grow much faster compared to the Au-rich FCC, increasing the BCC crystallite size from 14 nm to 199 nm, while the fcc crystals only grow up to 77 nm (Fig. 3B). This tendency agrees well with the crystallite size calculations for monometallic Au-fcc and Fe-bcc NPs during their heating to 700 °C (Fig. S3 in SI), where the size has doubled for Au (from 11 to 19 for smaller sized and 85 to 162 nm for larger sized) and increased 15 times for Fe (from 18 to 280 nm) NPs. A slight decrease in the crystallite size was observed by further heating, as an indication of the started evaporation of NPs and as already shown for other Au-Fe NPs. The temperature-induced recrystallization was also accompanied by the relaxation of $Au_{20}Fe_{80}$ NPs as quantified by microstrain for both FCC and BCC (Fig. 1E). Despite the weak BCC reflections at low temperatures (Fig. S1 in SI), microstrain could still be estimated (dotted line in Fig. 1E), indicating a more prestressed BCC compared to the FCC lattice. Considering their similar crystallite sizes up to 400 °C, such a difference in the microstrain can be attributed to more complex slip systems in bcc compared to fcc, originating from the different atomic packing factors (0.68 for bcc and 0.74 for fcc). [69, 70] A decreasing microstrain for both BCC/FCC at similar temperature region ($T_2 \approx 450$ °C) agreed with other Au-Fe compositions (Fig. 1C-E). *In situ* STEM analysis of the $Au_{20}Fe_{80}$ NPs match the XRD observations, providing further support to the temperature-dependent recrystallization and enhanced elemental mixing (Fig. 3C). It should be stated that unlimited homogenization would be, however, thermodynamically impossible even et elevated temperatures due to the limited miscibility of Au and Fe atoms in the bcc and fcc lattices, respectively. This behavior is well visible in the NPs with darker Fe-matrix containing lighter Au-nests as Fe-bcc can incorporate only 2-3 at.% Au. [30, 59] However, characterizing such segregations (nests) at the nanometer scale is highly challenging, especially in 3D. [71]

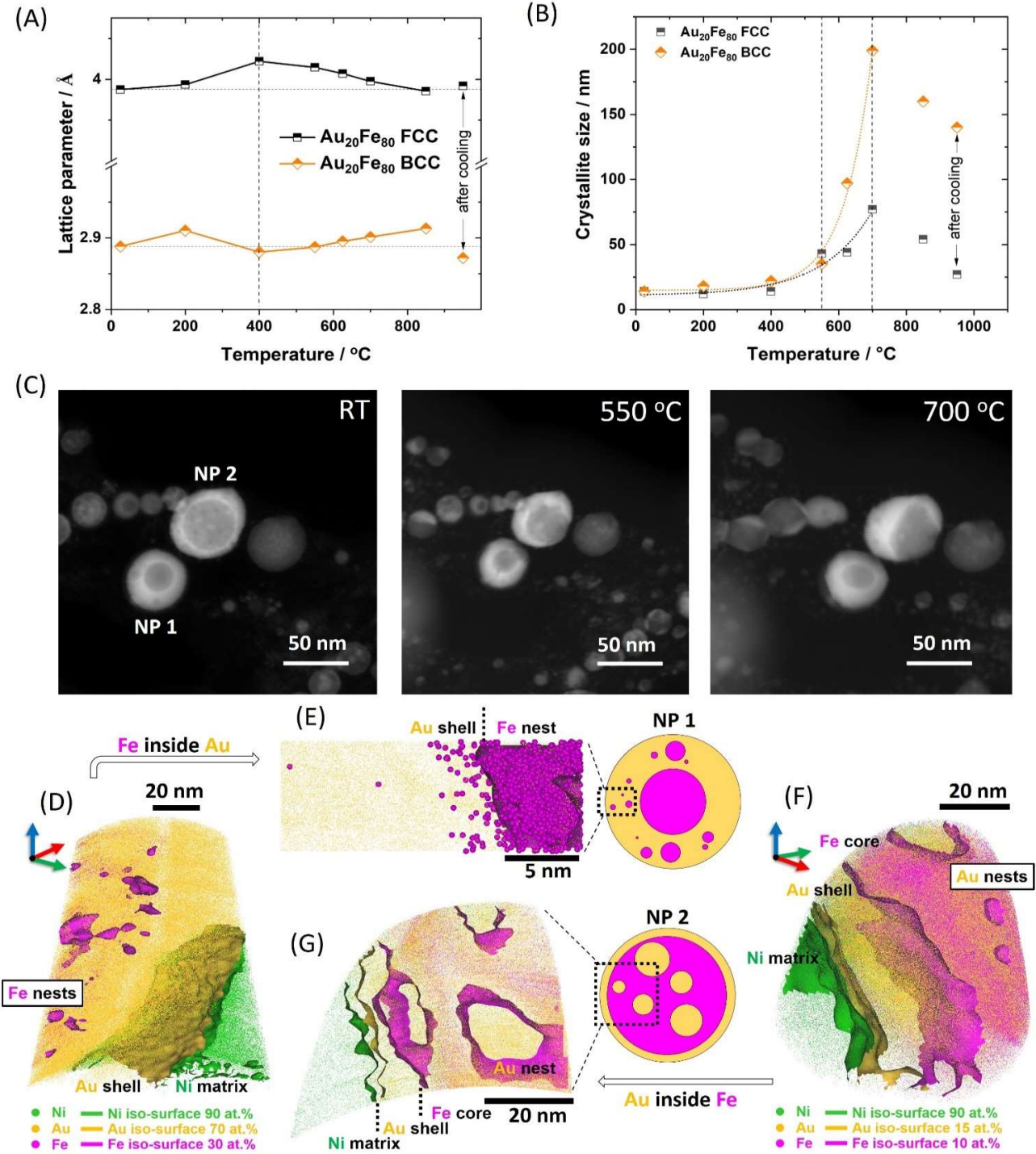

**Figure 3.** XRD derived (A) lattice parameters and (B) crystallite sizes of FCC and BCC phases for $Au_{20}Fe_{80}$ NPs (top). (C) Representative *in situ* STEM images of multiple $Au_{20}Fe_{80}$ NPs of at room temperature (RT), 550 °C and 700 °C with indicated two types of NP structures, e.g. NP 1 (Fe-nests inside Au-shell) and NP 2 (Au-nests inside Fe-core). (D, F) Reconstructed 3D atom maps from the acquired atom probe tomography (APT) dataset (C-related atoms are not shown) with (E, G) tomograms extracted from D and F, respectively. Each schematic inset image represents a corresponding nested core-shell NP 1 and NP 2 nanoparticle model.

As an advanced step, we hence performed APT analysis on $Au_{20}Fe_{80}$ NPs to enable a deeper internal view (Fig. 3D-F, S5, and S6 in SI). It should be noted here that the difference in the evaporation field between Au, Fe, and Ni elements could lead to aberrations in the ion trajectories that affect the shape of the reconstructed NPs.[72] For the selection of suitable 3D atom maps, two specific regions were selected and extracted to probe the existence of Fe-nests inside Au-shell (NP 1) and Au-nests inside Fe-core (NP 2). In the case of NP 1, many Fe-nests were found with an average size of about 4.6±3.1 nm (Fig. 3D, E). The selected iso-surface in NP 1 mainly represents the Au-shell (Fig. 3D) with the determined atomic ratio Au/Fe = 70/30, which agrees well with the FCC composition inferred from XRD. In the case of NP 2 with outer Au-shell and inner Fe-core, near-spherical Au-nest features of larger dimensions (up to 15 nm) are revealed (Fig. 3F, G). Here, the determined compositions at the iso-surface reflect both Au- and Fe-regions (Fig. 3F) in a comparable manner, so that the values (Au/Fe = 15/10) denote neither the Au-shell nor the Fe-core, but only represent the interfaces between the different phases.

To get insight into the possible formation mechanisms of the Au-Fe NPs and the respective roles of kinetics and thermodynamics, atomistic simulations were performed on annealing mixtures of Au and Fe. The molecular dynamics (MD) simulations were carried out by stepwise cooling of an initially hot but confined gas of 5000 atoms (corresponding to ~ 4.5 nm diameter) at the $Au_{50}Fe_{50}$ and $Au_{20}Fe_{80}$ compositions. A multishell pattern is always obtained, with a mixed core surrounded by a pure Fe layer and a full ($Au_{50}Fe_{50}$) or patchy ($Au_{20}Fe_{80}$) outer Au shell at 300 K and 7 ns annealing process (Fig. 4A). The mixed core is relatively ordered with clear intermetallic domains, and the outer Au and Fe surfaces exhibit crystalline features of fcc structure. The onion-shell pattern surrounded by pure shells is a very robust feature obtained for all trajectories and each composition, as well as simulations performed with only 2000 atoms (Fig. S7 and S8 in SI). Interestingly, at equilibrium, the present model used in the simulation does not favor mixed structures but elemental segregation.[73] The obtained inhomogeneous onion-shell pattern is thus a likely result of non-equilibrium effects of kinetic metastability. In agreement with the experimental findings which indicate inhomogeneities inside Au-Fe NPs, the simulated structures also show differently substituted areas in both compositions. Moreover, the MD simulations suggest that such inhomogeneities might appear even at very small scales down to 1 nm.

Since the present many-body potential used for the Au-Fe system does not particularly favor mixing between these elements,[73] further possible roles of kinetic versus thermodynamic stability in generating the yielded structures were investigated. Earlier parallel tempering Monte Carlo simulations did not find structures more stable in energy than CS configurations with pure Fe (bcc) in the cubic core coated by Au (fcc) in epitaxial contact.[73] As these NPs do not have the same sizes, it is not possible to directly compare the energies of these structures with the current MD ones. However, extrapolations can be attempted from a series of regular sizes and at the same composition. We note in passing that direct global optimization is out of reach for such rather large systems. From ideal models of nanocrystalline CS particles with a cubic bcc iron core and a sufficiently larger outer fcc gold shell in epitaxial contact, extra gold atoms were removed to reach the desired $Au_{50}Fe_{50}$ and $Au_{20}Fe_{80}$ compositions. At both compositions, the trends of Fig. 4B clearly show that the CS NPs are markedly lower in energy than the inhomogeneous onion-shell particles obtained upon annealing the metallic vapor. This indicates that the onion-shell structures would transform to more thermodynamically stable ones forming segregated phases (ideally, CS NPs) if given enough time and temperature.

To reproduce the transformation of the Au-Fe NPs found in *in situ* XRD and STEM, additional MD simulations were performed by reheating the NPs to 1000 K and cooling them back to 300 K after an intermediate step at 500 K (each temperature being maintained for 1 ns). It is noted that the morphology does not change significantly, although the extent of crystallinity and local ordering increases (Figs. S9, S10, and S11 in SI). At composition $Au_{20}Fe_{80}$, the occasional Au patches lying outside the NP sometimes coalesce or migrate towards the core, but in general, most of them remain at the surface (Fig. S9 in SI). Following the work by Doye and coworkers,[74] the strain energy of a locally minimized nanostructure was evaluated by calculating its many-body energy in the embedded atom model (EAM), but assuming nearest-neighbor pairs lie at the corresponding equilibrium distance. According to the calculations, $Au_{50}Fe_{50}$ and $Au_{20}Fe_{80}$ are initially characterized by 8 and 7 % strain, which decreases to 6 and 5 %, respectively, upon further heating, thus confirming the experimental relaxation trends discussed above.

While onion-ring structures were reported as the putative global minima of nanoalloys,[75] kinetics is generally found to play an essential role in their formation.[76] In agreement with our results, the onion-shell formation was also found for Ag-Au,[77] Ni-Pt,[78] and Mo-Cu[79] NPs as a result of kinetic trapping and the interplay between the (fast) coalescence of the core and the (slower) diffusion of elements within the NP. The present simulations additionally show an influence of temperature on the Au and Fe radial densities (Fig. 4C). Besides, this also helps to understand the formation mechanism of Au-Fe NPs in PLAL, evidencing the generation of a mixed core at very early stages for both compositions. After following the same Fe condensation into a pure shell on the formed core, differences in advanced steps are noticed (Figs. 4A, B). While $Au_{20}Fe_{80}$ NPs remain in the simple CS structure (residual Au atoms will condense later to form patches), Au depletion at intermediate distances occurs in $Au_{50}Fe_{50}$ and further increases during freezing and crystallization, leading to the formation of an extra pure gold outer shell. Upon further annealing, the crystalline feature becomes more pronounced, especially in the Fe-rich case (see also Figs. S7-S9 in SI).

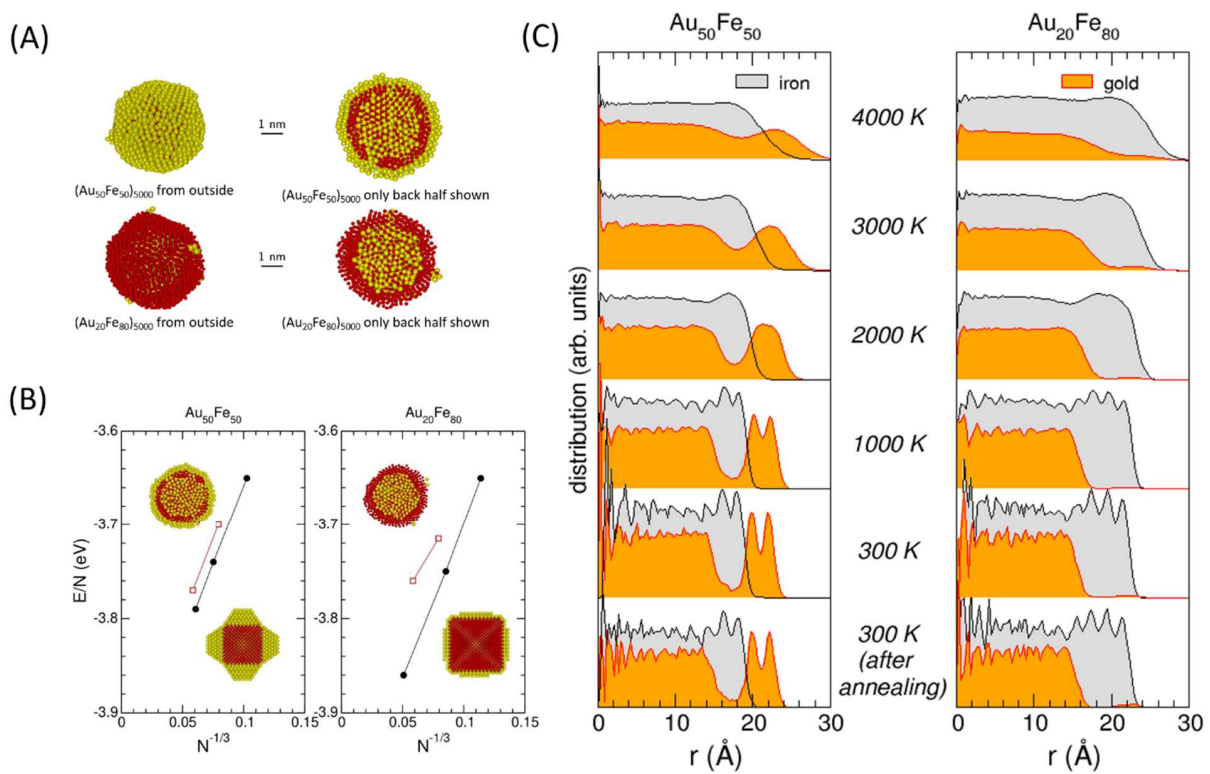

**Figure 4**. (A) Typical nanostructures obtained for 5000 atoms of gold and iron at the end of molecular dynamics trajectories of 7 ns long, at the $Au_{50}Fe_{50}$ and $Au_{20}Fe_{80}$ compositions. The left panels show the entire particles, while the right panels show truncated back halves. Iron and gold atoms are shown as red and yellow spheres, respectively (B) Binding energy per atom of Fe(core):Au(shell) nanoparticles (solid black circles) as a function of $N^{-1/3}$ where N is the number of atoms and compared with the average energies of the onion-shell nanoparticles obtained by annealing a hot vapor (hollow red squares), at the compositions $Au_{50}Fe_{50}$ (left panel) and $Au_{20}Fe_{80}$ (right panel). Iron and gold atoms are shown as red and yellow spheres, respectively (C) Radial distribution functions for gold and iron atoms averaged over 5 independent MD trajectories, at the compositions $Au_{50}Fe_{50}$ (left panel) and $Au_{20}Fe_{80}$ (right panel), as a decreasing function of temperature, from top to bottom. The corresponding functions after the additional annealing period are also shown.

Comparing now with the experimental results where the Fe-bcc nests were distributed in the Au-Fe fcc matrix, the simulated structures are constituted by Au-Fe fcc core surrounded by a Fe-bcc layer (and additional outer Au layer in $Au_{50}Fe_{50}$). Here it is important to note that our atomistic modeling was carried on much smaller NPs at extremely short time scales (ns) in contrast to the *in situ* XRD and STEM analyses.

Moreover, the formation of a mixed core seems rather universal in the gas phase, and the discrepancy between the measurements and the MD could well originate from the complete neglect of the liquid environment as well in our modeling. However, the existence of inhomogeneous structures in the Au-Fe nanosystem is confirmed in both experimental and theoretical pathways. Based on these, the laser-ablated Au-Fe NPs, featuring nanoscale inhomogeneities, are prestressed independent of their nominal composition. These particles can subsequently reach a more stable state (towards thermodynamic equilibrium) characterized by a homogenized Au-Fe elemental distribution when heated. Such a distribution is achieved through temperature-induced diffusion occurring due to the interactions not only between differently substituted fcc phases but also between the fcc and bcc lattices, enormously depending on the molar ratios of Au and Fe.

**CONCLUSIONS**

In summary, we showed the existence of non-equilibrium phases in laser-generated Au-Fe NPs characterized by the nanosized nests (phase segregations) inside individual NP components. Depending on the bulk composition and internal NP arrangements, the Au-Fe NPs crystallize into different fcc/bcc lattices with notable size-dependent microstrains. The range of Au-Fe compositions used in this study enabled a well distinguishable fcc-fcc ($Au_{80}Fe_{20}$), fcc-bcc/fcc ($Au_{50}Fe_{50}$) and fcc-bcc ($Au_{20}Fe_{80}$) interactions. The multiple fcc phases differing in substitution, crystallite size, and lattice strain, observed at moderate and Au-rich compositions, merge to single fcc phases at elevated temperatures through a facilitated diffusion-controlled atomic motion. We have thus demonstrated the multidimensional transformation of the initially stressed nest features in the Au-Fe NPs towards a more relaxed equilibrium state by applying sufficient thermal energy. Complementing the experimental observations made at the near-atomic scale, our MD simulations indicated the formation of a complex onion-like morphology having a mixed Au-Fe core, a Fe intermediate layer, and a Au-rich shell. The simulations with additional annealing steps also pointed towards the relaxation of a prestressed metastable system towards thermodynamic equilibrium, which we experimentally verified. These findings could constitute a key for fine-tuning the potential Au-Fe nanosystem for their application, e.g., in heterogeneous catalysis, where the temporal evolution of surface or internal segregations and associated crystal structures are of utmost importance and, in particular, a detailed understanding of their behavior at elevated process temperatures is highly relevant.

## MATERIALS AND METHODS

**Au-Fe NPs synthesis.** Au-Fe bimetallic NPs were synthesized by laser ablation of alloy targets with varying nominal compositions ($Au_{80}Fe_{20}$, $Au_{50}Fe_{50}$, and $Au_{20}Fe_{80}$, which correspond to the following mass ratios Au/Fe=93/07, Au/Fe=78/22 and Au/Fe=47/53) in acetone. A nanosecond pulsed laser (Rofin Sinar Technologies, Plymouth) with a wavelength 1064 nm, a repetition rate of 15 kHz, and pulse width 8 ns was employed for the ablation. In a typical experiment, the batch chamber was filled with 40 ml of acetone (Fisher chemical, 99.8%) in which the corresponding Au-Fe bimetallic target (fem Institut für Edelmetalle und Metallchemie, Schwäbisch Gmünd) was placed and then laser-irradiated horizontally for 15 minutes. The laser beam was scanned spirally over the surface of the target with the help of a beam scanner of focal length 10 mm, and liquid stirring was used for enhanced particle productivity. Monometallic Au and Fe NPs were also generated in the same way by ablating pure Au, and Fe foils in acetone for comparison. The as-generated bimetallic, as well as monometallic NPs, were having colloidal concentrations in the range 100-300 μg/mL.

***In situ* XRD analysis.** *In situ* X-ray powder diffraction was carried out to investigate the temperature-induced phase transition and recrystallization inside of Au-Fe NPs. For this study, a Panalytical Empyrean instrument in Bragg-Brentano mode with Cu Kα radiation (λ=1.54 Å; U=40 kV, I=40 mA), equipped with a high-temperature chamber HTK 16 (Anton Paar) was used. The Au-Fe NPs were homogeneously co-mixed with about 15 wt.% microcrystalline $LaB_6$ (standard reference material SRM 660b of NIST, a=4.15689 Å) and placed on the prestressed tantalum (Ta) strip heater ensuring high sample position stability. The standard enabled the instrumental correction and verification of the sample displacement at different temperatures needed in the Rietveld refinement. All samples were investigated in high vacuum ($10^{-3}$ Pa) at different temperatures (25 up to 850 °C) from 15 to 85° 2Θ with a step size of 0.01° so that each diffractogram was collected in 55 min. The heating rate between the individual isothermal steps was 10 K min$^{-1}$. After attaining the desired value, the temperature was kept constant for 10 min for thermal equilibration before starting the XRD measurement. Lattice parameters, phase amounts, crystallite sizes (determined using the Scherrer equation [58]), and microstrain (determined using the Stokes and Wilson equation [57]) of substituted Au-Fe NPs with fcc and bcc structure were determined by Rietveld refinement performed with the program package TOPAS 5 from Bruker using the fundamental parameter approach (FPA) and Pearson VII function. The achieved weighted profile R-factors ($R_{wp}$) of all Rietveld refinements varied between 4-5, confirming the reliable quality of the fits. [80] To investigate the temperature-induced relaxation (decreasing microstrain) in detail, the Williamson-Hall (W-H) plot technique [81] was employed. For the qualitative phase analysis with a Diffrac.Suite program EVA V1 from Bruker, the patterns of pure Au fcc (#4-0784) and Fe bcc (#6-0696), $LaB_6$ (#34-0427) reference, and Ta (#4-0788) sample holder from the ICDD database were used as references.

**Temperature-dependent STEM analysis.** The temperature-induced structural changes of NPs were investigated in the high-angle annular dark-field (HAADF) STEM mode, which allows atom number dependent Z-contrast. *In situ* STEM heating experiments were performed with a Gatan 652 double-tilt heating holder. Previous studies showed a stable temperature measurement. [82, 83] A heating rate of 10 °C per minute was applied, and the temperature was maintained for 10 minutes before the images were taken for each temperature plateau to minimize thermal drift and ensure a robust estimation of the sample temperature.

Heating experiments were performed with two samples: one sample with the nominal composition of $Au_{50}Fe_{50}$, generated in acetone with a picosecond laser (a 10 ps (Ekspla) Nd:YAG laser at 1064 nm with a repetition rate of 100 kHz and a fluence of 3.1mJ/cm$^2$) and a second sample with a nominal composition of $A_{20}Fe_{80}$, generated in 3-pentanone with the nanosecond laser (8 ns Nd:YAG laser (RofinSinar Technologies, Plymouth) at 1064 nm with a repetition rate of 15 kHz and a fluence of 3.85 mJ/cm$^2$).

All samples were investigated in a Tecnai F30 STwin G$^2$ TEM with 300 kV acceleration voltage. The colloidal samples were drop-coated on molybdenum TEM-grids (Plano GmbH) with lacy carbon carrier film. The elemental quantification was achieved using an energy-dispersive X-ray (EDX) system equipped with a Si/Li detector (EDAX system). For the cross-sectional sample imaging, the NPs were first embedded in a carbon matrix by a carbon injector followed by the focused ion beam (FIB) milling with a FEI Helios Nanolab system using a lift-out method.

**Atom probe tomography.** To fabricate APT specimens, freestanding nanoparticles first need to be encapsulated. The as-synthesized Au-Fe NPs were co-electrodeposited with Ni according to the protocol detailed in Ref. [84] The dried Au-Fe NPs powder was dispersed in a Ni ion electrolyte followed by pouring the solution into a vertical electro-cell where a counter electrode (Pt mesh) was positioned on top and a working electrode (Cu foil) placed on the bottom. Then, a low constant current (-19 mA) was applied to the electrodes to co-electrodeposit Ni and NPs on the working electrode. After the deposition, the composite film was kept in a vacuum desiccator.

The electroplated composite film was loaded into the focused ion beam (FIB) (Helios 600), and needle-shaped specimens were prepared following the *in-situ* lift-out protocol outlined in Ref. [85] APT measurements were carried out using a UV-laser assisted LEAP$^{TM}$ 5000 HR system (CAMECA Instruments Inc.) at an analysis stage temperature of 60 K. The laser repetition rate was set to 125 kHz, and the laser pulse energy was 60 pJ. Data was acquired with a detection rate of 5 ions for 1000 pulses on average. Datasets containing over 10 million ions were collected. Data analyses were performed using commercial software, IVAS 3.8.4.

**Atomistic simulation of Au-Fe nanostructures.** MD simulations were undertaken to investigate the possible structures of Au-Fe NPs formed upon laser ablation processes. We use a recently proposed many-body potential [73] parametrized for the Au-Fe system and based on original components for the pure elements of Au [86] and Fe. [87] In particular, the Fe potential correctly predicts the bcc structure as the most stable crystalline phase. [87] For the Au-Fe NPs, parallel tempering Monte Carlo simulations suggest that CS structures with a pure Fe core and pure Au shell are most stable with this model. [73] Here, we did not assume the structure of the nanostructures but simply simulated their formation by slow annealing from the hot vapor typically generated by laser ablation. More precisely, we placed a random distribution of atoms at the desired composition in Au-Fe inside a soft spherical container with radius $R$ creating a repulsive potential of $V(r)=k(r-R)^4$ for particle radii $r>R$ with $k=10^{-3}$ eV.Å$^{-4}$. Classical molecular dynamics simulations were then conducted at fixed temperature $T$, the thermostat being imposed by the Nosé-Hoover technique. After a fixed duration of 1 ns, the temperature was decreased, and the simulation continued. Here we should emphasize that the thermostat plays the implicit role of the heat bath provided by the surrounding liquid in the experiment, even though no molecules interacting with the metal atoms are explicitly considered.

In practice, the two compositions $Au_{50}Fe_{50}$ and $Au_{20}Fe_{80}$ were considered at the total sizes of 2000 and 5000 atoms. The temperature was decreased by steps of 500 K from 3000 K down to 500 K, followed by one more step at 300 K. The simulations were repeated independently 10 times for size 2000, and 5 times for size 5000, to ensure the reproducibility of the results. This annealing protocol thus corresponds to total simulated times of 7 ns. At size 2000, additional trajectories were also performed by equilibrating the metal vapor at 5000 K instead of 3000 K but keeping the same cooling rate of 1 ns per 500 K step, hence for a total of 11 ns.

Various properties were monitored upon the annealing process or for the finally obtained nanostructures. The crystallization process of the particles was quantified by calculating the bond-orientational order parameter $Q_6$, [88] between Au-Au, Fe-Fe, and Au-Fe bond vectors. These quantities correlate with the amount of local chemical ordering between the corresponding elements and are useful to distinguish amorphous versus ordered arrangements or random alloys versus intermetallics.

The morphologies and chemical ordering patterns were characterized by the radial distribution functions of the two elements relative to the center of the NPs, averaged over the last part of the trajectory at 300 K, but the direct visual depiction was considered as well.

# AUTHOR INFORMATION

**Corresponding Author**

* *Stephan Barcikowski*  stephan.barcikowski@uni-due.de

**Author Contributions**

‡J. Johny and O. Prymak contributed equally

# ACKNOWLEDGEMENT

The authors thank the German Research Foundation (DFG) project (BA 3580/18-1, KI 1263/15-1). Dr. Magali Benoit is gratefully acknowledged for providing the initial configurations with cubic iron cores used in the atomistic simulations. SHK, AAE, and BG acknowledge the financial support from ERC-CoG-SHINE-771602.